\documentclass{mscs}

\usepackage{authordate1-4}
\usepackage{t1enc} 
\usepackage{theorem} 
\usepackage{amssymb} 
\usepackage{fancyvrb}
\usepackage[breaklinks]{hyperref}
\usepackage{color}
\usepackage{graphics}
\usepackage[all]{xy}

\DefineShortVerb{\|}

\newcommand{\comment}[1]{}

\newcommand{\hsp}[1][3ex]{\hspace*{#1}}

\newcommand{\eg}{{\em e.g.} }
\newcommand{\ie}{{\em i.e.} }

\newcommand{\etc}{{\em etc.}}
\renewcommand{\inf}{\infty}
\newcommand{\rencap}{\textsc{rencap}}

\renewcommand{\a}{\rightarrow}

\newcommand{\la}{\leftarrow}

\newcommand{\I}[1]{[\![#1]\!]}

\newcommand{\sle}{\subseteq}

\newcommand{\tge}{\unrhd}

\renewcommand{\prod}{_\mr{prod}}

\renewcommand{\b}{\beta}

\newcommand{\vep}{\varepsilon}

\renewcommand{\t}{\theta}

\renewcommand{\r}{\rho}
\newcommand{\s}{\sigma}
\renewcommand{\S}{\Sigma}

\newcommand{\mc}{\mathcal}

\newcommand{\mr}{\mathrm}
\newcommand{\mb}{\mathbb}

\newcommand{\bN}{\mb{N}}

\newcommand{\cG}{\mc{G}}

\newcommand{\cT}{\mc{T}}

\newcommand{\cX}{\mc{X}}

\newenvironment{rul}
  {$\begin{array}{rcl}}
  {\end{array}$}

\newenvironment{rew}[1][~~\a~~]
  {$\begin{array}{r@{#1}l}}
  {\end{array}$}
\newenvironment{rewc}[1][~~\a~~]
  {\begin{center}\begin{rew}[#1]}
  {\end{rew}\end{center}}

{\theorembodyfont{\rmfamily} 

}

\newenvironment{lstgeneric}[2]
  {\begin{list}{#1}{\topsep=.5mm\itemsep=.5mm\parsep=0mm%
    \itemindent=-3ex\labelsep=1ex\labelwidth=0ex #2}}
  {\end{list}}

\definecolor{darkgreen}{rgb}{.3,.7,.3}

\newcommand{\colbox}[3]{\textcolor{#1}{\textsf{\textbf{[#2:]}} #3}}
\renewcommand{\colbox}[3]{}

\newcommand{\adam}[1]{\colbox{blue}{Adam}{#1}}

\newcommand{\aint}{\a^{\hsp[-3.5mm]>\vep\hsp[-1mm]}}
\renewcommand{\atop}{\a^{\hsp[-2.5mm]\vep\hsp[-1mm]}}

\newcommand{\cime}{CiME3}
\newcommand{\ceta}{CeTA}
\newcommand{\isafor}{IsaFoR}

\begin{document}

\title[CoLoR: a Coq library on well-founded rewrite relations]
{CoLoR: a Coq library on well-founded rewrite relations
and its application to the automated verification of termination certificates}

\author[F. Blanqui and A. Koprowski]
{F\ls R\ls \'E\ls D\ls \'E\ls R\ls I\ls C\ns B\ls L\ls A\ls N\ls Q\ls U\ls I$^1$ 
\ns and\ns 
 A\ls D\ls A\ls M\ns K\ls O\ls P\ls R\ls O\ls W\ls S\ls K\ls I$^{2}$\\
$^1$ INRIA, FIT 3-604, Tsinghua University, Haidian District, Beijing
100084, China
\addressbreak
$^2$ MLstate, 15 Rue Berlier, 75013 Paris, France}

\maketitle

\begin{abstract}
Termination is an important property of programs; notably required
for programs formulated in proof assistants. It is a very active 
subject of research in the Turing-complete formalism of term rewriting.
Over the years many methods and tools have been developed to address
the problem of deciding termination for specific problems (since it is
undecidable in general). Ensuring reliability of those tools is 
therefore an important issue.

In this paper we present a library formalizing important results 
of the theory of well-founded (rewrite) relations in the proof assistant 
Coq. We also present its application to the automated verification 
of termination certificates, as produced by termination tools.

The sources are freely available at \url{http://color.inria.fr/}.
\end{abstract}

\adam{If you want to get rid of \textrm{adam/frederic/todo} comments
I made it so that it is enough to uncomment line 4 in \textrm{macros2.tex}.
So to prepare submission/draft versions I think it's better to do that than
to remove all the comments.}

\section{Introduction}
\label{sec-intro}

Rewriting is a general (Turing-complete) yet very simple formalism
\cite{
terese03book} that can be used as
a programming language \cite{
elan,maude} or in which some
other programming languages can be easily encoded (in particular logic
and functional programming languages). Both cases open the way to
benefit from techniques
developed for term rewrite systems like termination
\cite{
giesl06rta,nguyen07lopstr,schneiderkamp09tocl} or
complexity analysis \cite{marion03ic,hirokawa08ijcar}. Term rewriting
is also a general tool for deciding the equality of two terms in 
some equational theory
\cite{knuth67}.

That is why various authors proposed logical systems where functions
and predicates can be defined by arbitrary user-defined rewrite rules
(instead of inductive definitions only), and where the equivalence on
types/propositions is enriched with these user-defined rules
\cite{coquand92types,barbanera97jfp,dowek03jar,blanqui05mscs}.
This is especially important, as
enriching the equivalence on types facilitates
the use of dependent types. However, in contrast with the systems
where all functions and predicates are inductively defined
\cite{coquand88colog,altenkirch93phd,werner94phd}, the decidability of
type-checking and the logical consistency of the system are not
guaranteed anymore. To ensure these essential properties, the
user-defined rules must satisfy non-trivial conditions like subject
reduction, confluence, termination or definition completeness
\cite{coquand92types,dowek03jsl,blanqui05mscs,walukiewicz08lmcs}. Checking
such conditions requires the use of complex and thus likely to be buggy
software, which reduces the overall confidence we may have in such
logical systems.

Coq is a proof assistant based on the calculus of inductive
constructions \cite{
coquand88colog,werner94phd}, a very
rich typed lambda-calculus with polymorphic and dependent (inductive)
types, that includes higher-order logic through the
propositions-as-types principle
\cite{
barendregt92book}. For a complete overview
of this proof assistant, we refer the reader to the Coq reference
manual \cite{coq} and to \cite{bertot04book}; instead we only mention
some of its features: functions and predicates
can be defined inductively \cite{paulin93tlca}, proof terms are
obtained by executing scripts \cite{delahaye00lpar}, definitions and
proof scripts can be organized in modules \cite{chrzaszcz03tphol}.

A \emph{rewrite relation} is simply a relation on the set of
first-order terms generated from a given signature that is stable under
substitutions and contexts. In general, one considers rewrite
relations generated from a set of rewrite rules.
A rewrite relation is \emph{terminating} (strongly normalizing,
well-founded, noetherian) if there is no term starting an
infinite sequence of rewrite steps.

In this paper, we present the foundations of a formalization of the
theory of well-founded first-order rewrite relations
\cite{
terese03book} in Coq. 
There are various motivations for this work:

\begin{itemize}
\item
verifying correctness of certificates produced by automated
termination provers;
\item
allowing function definitions with
non-struc\-turally recursive calls in proof assistants, 
and the use of external automated
termination provers to check their termination;
\item
providing an important library of types and functions making an
extensive use of dependent types.
\end{itemize}
We will elaborate on them in the following sections.

\subsection{Verifying correctness of termination certificates}

In the last years, many new techniques and tools have been developed
to automatically (dis)prove termination of rewrite systems
\cite{tp,tc,waldmann09wst}. These techniques and tools are more and
more sophisticated, and use external tools like SAT solvers
\cite{schneiderkamp07frocos,fuhs07sat}. 
As a consequence, it is hard to trust their results and, indeed,
every year sees some tools disqualified
because of errors found in their results.

Hence, providing a way to automatically verify correctness of
termination certificates is useful for many applications like
automated termination provers and proof assistants.

\subsubsection{Termination certificates.}

Termination techniques can be divided into two categories: the ones that
either prove a complete termination problem or fail, and the ones that
transform/split a termination problem into one or more termination
problems. Hence, a termination proof can be represented by 
a termination certificate that, roughly speaking, is a tree, 
a node being labelled by a termination problem, the termination 
technique applied at that point and its parameters.

For instance, a rewrite system can be proved terminating by finding a
well-founded quasi-ordering on symbols so that, for each rule, the
left-hand side is strictly greater than the right-hand side in the
multiset path ordering (MPO) based on the quasi-ordering on symbols
\cite{dershowitz82tcs}. Then, a certificate for this proof can be
given by a node labelled with the termination problem, the termination
technique used (MPO) and its parameters (the quasi-ordering on symbols).

Verifying that a certificate is correct consists of checking that
parameters indeed satisfy the conditions required by the termination
technique and, in case of a transformation technique, that the
subtrees are labelled by correct termination problems. Checking
that some parameters satisfy the conditions required by a termination
technique should be of reasonable complexity (polynomial). Otherwise,
the certificate must be refined by providing more details on how to
check the conditions. In the case of MPO, given a finite well-founded
quasi-ordering on symbols, one only has to check that each rule is
included in MPO.

A common format based on these ideas has been recently developed and
used in the last termination competition \cite{cpf}.

\subsubsection{Our approach.}

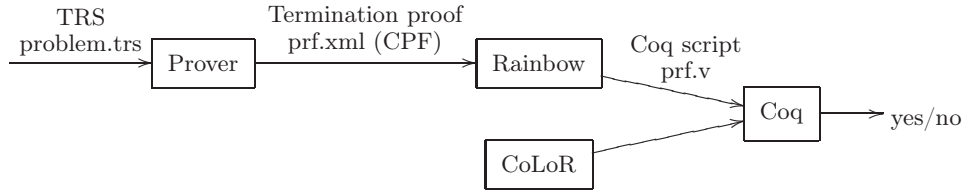
\begin{figure}
\xymatrix@R=0pt{
\ar[rr]^-{\txt{TRS \\ problem.trs}} &&
*++[F]\txt{Prover} \ar[rrr]^-{\txt{Termination proof \\ prf.xml (CPF)}} &&&
*++[F]\txt{Rainbow} \ar[rrd]^-{\txt{Coq script \\ prf.v}} && \\
&&&&&&& *++[F]\txt{Coq} \ar[r] & \txt{yes/no} \\
&&&&& *++[F]\txt{CoLoR} \ar[rru]
}
\vspace{0.2cm}
\caption{Certifying termination with CoLoR and Rainbow}
\label{fig-color}
\end{figure}

In order to verify such termination certificates with the highest
confidence possible, we developed the following methodology (see
Figure \ref{fig-color}).

We formalized in a proof assistant (Coq) various termination
techniques used in modern automated termination provers, together with
boolean functions for checking the conditions that the parameters must
satisfy, and we proved their correctness. This is the CoLoR library
that we are going to describe in the following sections.

Then, we wrote a program, called Rainbow, which, given a file
containing a certificate in the aforementioned format, generates a Coq
file containing a formalization of the termination problem, a
formalization of all the parameters used in the certificate, and a
short and simple proof script for the theorem stating that the rewrite
system is (non-)terminating. This script consists of applying the
lemmas corresponding to the termination techniques used in the
certificate, and checking correctness of the parameters by testing
the equality of the corresponding boolean functions to true
(reflexivity proof). Coq is then called to check the correctness of
this proof script.

We tried to keep Rainbow as simple as possible but, nonetheless,
Rainbow itself can introduce various types of errors: parsing can be
wrong or incomplete, and the formalization of the termination problem,
the parameters or the termination proof may be wrong. Moreover, some
termination proofs may require a lot of computations (for instance,
some proofs use dozens of successive matrix interpretations), and
computations in Coq are often much less efficient than in more traditional
programming languages.

To improve this, we started to formalize Rainbow itself in Coq by
formalizing the certificates themselves, defining boolean functions
to check certificate correctness and proving that this
function itself is correct, that is, that one can indeed build a
termination proof if the function returns true. Then, by using Coq
extraction mechanism
\cite{paulin89popl,letouzey02types}, 
we can get a
standalone termination certificate checker.

\subsubsection{Results.}

\comment{
In the last experiments, on the 1336 termination problems of the
termination problem database (TPDB) \cite{tpdb} consisting of standard
rewrite systems, Rainbow was able to successfully verify 
correctness of 701 certificates produced by the automated termination
prover AProVE \cite{giesl06ijcar} (best prover for this category in
last competitions). In the 2008 competition, AProVE could find a proof
for 1226 \comment{of the 1391} termination problems in this
category\comment{(the database and the problems submitted to the
  competition are evolving every year)}. Hence, currently Rainbow can
verify about 62\% of termination proofs produced by AProVE.}

In 2009, some experiments showed that Rainbow could successfully check
up to 701 certificates among 1226 produced by AProVE
\cite{giesl06ijcar}, that is, 57\%. In the 2009 termination
competition, it could successfully check 399 certificates among the
964 generated by the participating provers, that is, 41\%. (Rainbow
did not participate to the 2010 competition held only 6 months after
the 2009 competition.) This difference is due to the fact that, in the
first case, the certificates were produced specifically for Rainbow
while, in the second case, many certificates were produced for tools
handling termination techniques not supported by Rainbow. The best
tool was CeTA \cite{sternagel09tphol} with 774 certificates
successfully checked, that is, 80\%.

The project started in 2004 and since then the 
Rainbow program grew to about 6,000 lines of OCaml code
and the CoLoR library to approximately 70,000 lines of Coq, with its contents
roughly broken down as follows:
\begin{itemize}
 \item 12\%: basic mathematical theories (relations, semi-rings), 
 \item 25\%: data structures (lists, vectors, multisets, matrices,
   polynomials),
 \item 38\%: term structures and rewriting theory,
 \item 25\%: termination techniques.
\end{itemize}
It contains about 1500 definitions
(types, predicates or functions), and about 3500 theorems, many of
them being of course simple theorems stating introduction/elimination
rules for the defined notions, or equalities/equivalences used to
explicitly rewrite terms or propositions in proofs. The library
also provides about 185 (simple) tactics \cite{delahaye00lpar} that
are used in CoLoR or in the termination proof scripts generated by
Rainbow.

Developing libraries of functions and theorems on some data structures
or basic mathematical theories is interesting in itself since it can
be reused by other people for doing other developments. This is the
case for CoLoR which has so far been used in a formalization of
the Spi calculus \cite{briais08phd}, a modular development of
certified program verifiers \cite{chlipala06icfp} and an efficient
Coq tactic for deciding Kleene algebras \cite{braibant10itp}.

\subsection{Using dependent types}

Another motivation for this work was to make a development heavily
using dependent types
\cite{
barendregt92book}. First-order terms with
symbols of fixed arity can be naturally formalized using dependent
types as we will see in the next sections. However, it is well known that,
in current proof assistants, dependent types may sometimes be
difficult to work with, because the equivalence on types is not rich
enough. Such developments can therefore be used as a benchmark for
evaluating proof assistants on the feasibility and ease of use of 
dependent types. Currently,
to address this problem, one needs to introduce auxiliary functions or
explicitly apply type casting functions. As already mentioned, the use
of rewriting or (certified) decision procedures in the equivalence on
types, solves many of these problems
\cite{blanqui05mscs,blanqui08tcs,strub10csl}.
This approach is adopted
in Coq modulo theories (CoqMT), a new extension of Coq where type
equivalence can be extended by decision procedures such as linear
arithmetic \cite{coqmt}.

\bigskip

\subsection{Outline}

Preliminary overview of CoLoR was given in
\cite{blanqui06wst-color,koprowski08phd,blanqui09tr}. Detailed
descriptions of formalizations of some particular termination
techniques are presented in prior publications, to which we refer for
more details:

\begin{itemize}
 \item 
   polynomial interpretations
   \cite{hinderer04master},
 \item
   recursive path ordering
   \cite{coupet06aaecc},
 \item 
   multiset ordering and higher-order recursive path ordering
   \cite{koprowski06rta,koprowski08phd,koprowski09aaecc},
 \item
   matrix and arctic interpretations
   \cite{koprowski08sofsem,koprowski08rta,koprowski08phd}.
\end{itemize}

In this paper, we give a detailed presentation of CoLoR's foundations
(definitions of terms, rewriting, \etc) and of key termination
techniques not presented before (dependency pairs, dependency graph
decomposition and reduction pairs).

The remainder of the paper is organized as follows.

In Section \ref{sec-related}, we discuss related work.

In Section \ref{sec-term}, we present our formalization of the basic
notions of rewriting theory: signatures, terms, contexts,
interpretations, substitutions, rules, rewriting and termination. We
also discuss some of the problems we faced in Coq and how we addressed 
them.

In Section \ref{sec-dp}, we describe the formalization and the proof of
the main theorem on dependency pairs, a key notion of modern automated
termination provers. This development provides an interesting example of
a use of a higher-order, polymorphic and dependent program (for
computing the cap and aliens of a term).

In Section \ref{sec-graph}, we present the formalization of the
dependency graph decomposition, another key technique of modern
automated termination provers.

In Section \ref{sec-manna-ness}, we describe the formalization of the
general termination technique based on reduction pairs/orderings. A
particular instance of this technique is the polynomial
interpretations over natural numbers described in Section
\ref{sec-poly}.

In Section \ref{sec-example}, we illustrate our approach by presenting 
a complete example of automatically generated Coq script for some simple 
termination certificate.

Finally, Section \ref{sec-conclu} provides some concluding remarks and
presents future directions of research.

\section{Related work}
\label{sec-related}

There are two other libraries/tools aiming at verifying termination
certificates that participate in the termination competition
\cite{tc}: \cime~\cite{cime3} and \ceta~\cite{ceta}. In addition, proof
assistants generally have their own termination checkers.

\subsection{\cime}

\cime\ also produces Coq scripts based on a Coq library called Coccinelle
\cite{
contejean07jpbirthday,contejean07frocos,courtieu08tphol}. The
approach taken in that tool is slightly different from ours.
In Rainbow, we use a \emph{deep embedding}: every type of
objects and every termination technique used in certificates is
formalized in CoLoR and can be the subject of a mathematical study. In
\cime, this is not always the case: a \emph{shallow embedding} is used
for some types of objects and some termination techniques.

For instance, to formalize a termination problem, Rainbow defines the
set of rules and uses the definition of rewrite relation generated by
a set of rules defined in CoLoR. On the other hand, \cime\ generates
an \emph{ad hoc} inductive predicate defining the rewrite relation. Hence, the Coq
scripts generated by \cime\ are often longer, less readable and take
significantly more time to be checked by Coq.

The main advantage of the shallow embedding approach is the 
ability to leverage some features of the proof assistant 
to get some work done for free. For instance defining 
polynomials as native functions in the proof assistant
(shallow embedding), instead of as a new inductive data-type
(deep embedding), forbids us from studying their meta-theory but 
gives us for free the capability to evaluate a polynomial on 
given values.

However, because the amount and the complexity of 
the generated code is more important, we think that this makes 
\cime\ more difficult to develop and maintain. Even more importantly, 
the use of a shallow embedding makes it impossible to develop
meta-theory and hence generic checkers for termination techniques.
Therefore one cannot benefit from the Coq extraction mechanism
to obtain an independent, certified checker; an approach that we
plan to incorporate to our toolset in the near future.

Some notions or techniques can be found in both libraries, but they
are often formalized differently and work with different notions of
terms. Indeed, CoLoR and Coccinelle do not define terms in the same
way (we will come back to this point in the next section). However, in
CoLoR, we defined a translation of CoLoR terms into Coccinelle terms
in order to reuse some results/functions available in Coccinelle only,
like its certified decision procedures for
matching modulo AC (associativity and commutativity),
unification modulo ACU (associativity and commutativity with a neutral element)
and the recursive path ordering (RPO). Hence,
the last version of Rainbow could verify RPO proofs by using
Coccinelle. The converse translation (of Coccinelle terms to CoLoR
terms) should also be possible, although slightly more complicated,
as it would have to translate a Coccinelle term to an \emph{optional}
CoLoR term, checking term well-formedness in the process.

\subsection{\ceta}

\ceta\ is a Haskell \cite{haskell} program (there is also an OCaml version
\cite{ocaml}) extracted
\cite{berghofer02types,haftmann10flops} 
from a library
called \isafor~\cite{sternagel09tphol,sternagel10rta} 
developed in the proof assistant Isabelle/HOL
\cite{nipkow02lncs}. Hence, it also uses a full deep embedding
approach and is naturally faster than \cime\ and the current
non-extracted version of Rainbow.

\isafor\ now includes most techniques previously formalized in CoLoR and
Coccinelle and some new important ones (most notably the subterm criterion
and usable rules) that greatly increase the number of termination
proofs that can be verified. Apart from termination techniques,
also parsing of termination certificates is formalized in \isafor,
so that extraction gives a complete, stand-alone tool for checking
termination proofs. \isafor\ is now nearly 40,000 lines of Isabelle/HOL. 

At the moment, \ceta\ is the best termination certificate verifier: 
at the time of writing, it can successfully verify 1432 certificates 
over 1536 found by TTT2 \cite{korp09rta} on 2132 termination 
problems, hence it is capable of certifying 93\% of TTT2 proofs.

The main difference between \isafor\ on the one hand, and Coccinelle and
CoLoR on the other hand, is therefore the language and the axioms used
to define objects and properties. In \isafor, it is a
ML-like \cite{harper86tr} simply typed lambda-calculus with
(implicitly universally quantified) type variables and inductive
predicates, while in Coccinelle and CoLoR, the language used is the
calculus of inductive constructions
\cite{coquand88colog,werner94phd} featuring fully
polymorphic and dependent types. Moreover, in \isafor, the (higher-order) logic
imposed by the system is {\em classical},
while in Coccinelle and CoLoR it is {\em intuitionistic}.
It is however possible, and sometimes
necessary, to use in Coq the excluded middle 
(we will come back to this point later). Hence, in \isafor, termination
is defined as the absence of infinite sequences of rewrite steps,
while in Coccinelle and CoLoR, 
termination is defined as a constructive inductive predicate called
accessibility (more on that in Section~\ref{subsection:termination}). 
Since most of the correctness proofs of termination
techniques that one can find in the literature are classical, their
adaptation to a constructive setting is an interesting endeavour.
We leave for future work a more detailed comparison of the two approaches.

\subsection{Proof assistants}

Proof assistants such as Coq or Isabelle/HOL
include their own automated termination provers but these
provers are implemented as internal tools working on the
internal representation of the proof assistant language and,
unfortunately, currently can not be used outside these proof
assistants.

Coq has its own termination prover developed by Bruno Barras and
based on \cite{gimenez94types} that essentially checks that recursive
call arguments are ``structurally smaller'' \cite{coquand92types}
modulo some possible reductions. Coq also provides the |Function|
command \cite{balaa00tphol,barthe06flops} 
and the more
recent |Program| extension \cite{sozeau07icfp}, both of which allow
one to define a function with non-structurally decreasing arguments in
recursive calls, provided that one can prove that all
recursive call arguments are strictly decreasing in some well-founded
relation.

Isabelle/HOL has its own automated termination provers that internally
generate Isabelle/HOL termination proofs
\cite{bulwahn07tphol,krauss07cade}. 

Since those embedded termination provers are necessarily limited,
proof assistants would greatly benefit from the certification of
termination proofs allowed by \ceta, \cime\ and Rainbow.

\section{Terms and rewrite relations}
\label{sec-term}

In this section, we explain how terms and rewriting concepts are 
formalized in CoLoR. We encourage the reader to consult the 
actual sources of CoLoR for a more in-depth understanding of the
presented notions. The sources are available for download
and for online browsing (with proofs omitted) at:
\begin{center}
 \url{http://color.inria.fr/}.
\end{center}

CoLoR provides various notions of terms: strings, first-order terms
with symbols of fixed arity (simply called algebraic in the
following), first-order terms with varyadic symbols (a varyadic symbol
can be applied to any number of arguments; this is often used when a
symbol is associative and commutative), and simply typed lambda terms
\cite{koprowski09aaecc}. The files about strings are prefixed by |S|,
those about varyadic terms are prefixed by |V|, and those about
algebraic terms are prefixed by |A|.

In this paper, we will only consider algebraic terms (directory
|Term/WithArity|).

\subsection{Signatures}

Algebraic terms are inductively defined from a signature (module
|ASignature| defined in file |ASignature.v|) defining the set of
symbols, the arity of each symbol, a boolean function saying if two
symbols are equal or not, and a proof that this function is correct
and complete wrt Coq default (Leibniz) equality predicate:

\begin{Verbatim}[fontsize=\small]
Record Signature : Type := mkSignature {
  symbol :> Type;
  arity : symbol -> nat;
  beq_symb : symbol -> symbol -> bool;
  beq_symb_ok : forall x y, beq_symb x y = true <-> x = y }.
\end{Verbatim}

Note that we declare the record field selection function |symbol:|
|Signature| |-> Type| as an implicit coercion (keyword |:>|)
\cite{saibi97popl} 
so that a signature can always be given
as argument to a function waiting for a |Type|, the Coq system adding
the necessary calls to |symbol| wherever necessary.

Another solution would be to make |symbol| a parameter of a
signature. It would add yet another argument to functions and lemmas
using a signature, but this argument can be inferred by the Coq system
and the extracted OCaml code would be cleaner and easier to use (no
|Obj.magic|). It would also allow us to use the new Coq feature of
type classes \cite{sozeau08tphol}. We leave for future work such an
experimentation.

At the beginning, we were using a single lemma stating the decidability of
equality on the set of symbols:
\begin{center}|eq_dec: forall x y, {x=y}+{~x=y}|,\end{center}
where |{A} + {B}| is the standard Coq notation for disjunction between
|A| and |B| (the difference with the disjunction notation |A \/ B| having
to do with extraction).
Separating into a computational function (|beq_symb|) and a proof
of its correctness (|beq_symb_ok|), as used, for instance, in the
Ssreflect library \cite{gonthier09tr}, improved the
time spent by Coq for checking scripts generated by Rainbow
by 9\% on average. This can
be explained by the fact that terms are smaller and fewer conversion
tests are required.

\subsection{Vectors and equality}

For defining terms, we use the Coq module |Bvector|\footnote{The ``B''
  is there for historical reasons, as this library started
  as a library of vectors over booleans.} of the standard library
which defines the type of vectors (also called arrays or dependent
lists) with elements of type |A|:

\begin{Verbatim}[fontsize=\small]
Inductive vector : nat -> Type :=
  | Vnil : vector 0
  | Vcons : forall (a:A) (n:nat), vector n -> vector (S n).
\end{Verbatim}

But the Coq standard library provides almost no functions and
theorems about vectors. We therefore had to develop an important
library on vectors (directory |Util/Vector| and, in particular, the
module |VecUtil|). To our knowledge, this is the most developed
library on vectors (more than 3,000 lines of Coq).

A simple yet important function that we need sometimes to use for
fulfilling some type constraints is the following explicit type
casting function which, given a vector of size |m| (|vec| is an
abbreviation for |vector A|) and a proof that |m=n|, returns a vector
of size |n|:

\begin{Verbatim}[fontsize=\small]
Fixpoint Vcast m (v : vec m) n (mn : m = n) : vec n := ...
\end{Verbatim}

This function is of course nothing but the identity from a
computational point of view, but it allows one to see a vector of size
|m| as a vector of size |n| whenever |m| and |n| are
provably equal. It is currently defined as a recursive function
breaking up and building back the vector. But it is also possible to define it
as the identity (experimentation done by Pierre-Yves Strub but not integrated
in CoLoR yet).
It is needed in some theorems which would not be typable otherwise,
such as:

\begin{Verbatim}[fontsize=\small]
Lemma Vapp_nil : forall n (v : vec n) (w : vec 0), 
  Vapp v w = Vcast v (plus_n_O n).
\end{Verbatim}

\noindent
since |Vapp v w| is of type |n+0| which is not computationally but
only inductively equivalent to |n|, because the addition is defined by
induction on its first argument. Indeed, in Coq, equality is defined
as follows:

\begin{Verbatim}[fontsize=\small]
Inductive eq (A:Type) (x:A) : A -> Prop := refl_equal : eq A x x.
\end{Verbatim}

It is possible to avoid these explicit casts by instead using Conor
McBride's inductive definition of equality \cite{mcbride99phd} which,
to some extent, allows both members of the equality to have distinct
types:

\begin{Verbatim}[fontsize=\small]
Inductive JMeq (A:Type) (x:A) : forall B:Type, B -> Prop := JMeq_refl : JMeq x x.
\end{Verbatim}

But, for eliminating such an equality, both types must be
computationally equivalent. So, we prefered to stick with the standard
equality of Coq, especially since tactics available in Coq are better
suited for reasoning about it.

Note also that, in some cases, we need to use the identity of equality
proofs (|forall| |h1 h2| |:| |n=m|, |h1 = h2|)
\cite{streicher93hdr}. Since the equalities that we consider are on
the set |nat| of natural numbers on which equality is decidable, this
property can be proved and no axiom is required.

We could perhaps benefit from a recent library of lemmas and tactics
to reason on heterogeneous equalities \cite{heq}. But we will prefer
to use an extension of Coq incorporating rewriting or decision
procedures in the equivalence of types
\cite{blanqui05mscs,blanqui08tcs,strub10csl}. Some recent experiments
show that, with CoqMT \cite{coqmt}, all casts can be
removed\footnote{See
  \url{http://git.strub.nu/git/coqmt/tree/test-suite/dp/dlist.v}.}.

\subsection{Terms}

Given a signature |Sig|, we can define the type of algebraic
terms (module |ATerm|):

\begin{Verbatim}[fontsize=\small]
Notation variable := nat (only parsing).
Inductive term : Type :=
  | Var : variable -> term
  | Fun : forall f : Sig, vector term (arity f) -> term.
\end{Verbatim}

Note that variables are represented by natural numbers. It is
sufficient and simpler than using an arbitrary type because such a
type needs to be equipped with functions and properties for, say,
defining the renaming of a term away from some finite set of
variables. By using natural numbers, we directly benefit from all the
functions, properties and tactics available on natural numbers.

Thanks to the strong type system of Coq, every expression of type
|term| corresponds to an algebraic term in the mathematical sense: a
symbol cannot be applied to a number of terms distinct from its arity.

Another solution would be, like it is the case in Coccinelle, to
define terms as varyadic terms, together with a function for checking
the well-formedness of a term that one would use whenever it is
necessary:

\begin{Verbatim}[fontsize=\small]
Inductive term : Type :=
  | Var : variable -> term
  | Term : symbol -> list term -> term.
Fixpoint well_formed (t:term) : bool := ...
\end{Verbatim}

The advantage is that various termination criteria defined in the
literature for algebraic terms are in fact also valid for varyadic
terms. However, when one needs to reason on well-formed terms, it is
much more convenient to use an inductive definition that provides
these conditions for free. We will see such an example with polynomial
interpretations (Section \ref{sec-poly}) which is a technique that
naturally requires well-formed terms, since every symbol of arity $n$
must be associated to a polynomial with $n$ variables.

Since algebraic terms can easily be translated into varyadic terms,
every termination criterion developed for varyadic terms can be easily
applied on algebraic terms. Hence, by using a translation from CoLoR
algebraic terms to Coccinelle varyadic terms (module |Coccinelle|), we
could enable Rainbow to verify certificates for RPO by reusing the
efficient decision procedure for RPO developed in Coccinelle.

It is certainly possible to define a translation the other way around
and, for \cime, to reuse results formalized in CoLoR on algebraic
terms. As remarked before, such a translation would transform
a Coccinelle term into an \emph{optional} CoLoR term, while checking
term well-formedness along the way.

\subsubsection{Remarks on recursive definitions in Coq.}

The induction principle automatically generated by Coq for the type
|term| is too weak since |vector| |term| |(arity f)| or |list| |term|
are not seen as recursive arguments. We therefore need to redefine it
by hand.

Because of the limitation of Coq termination checker, it is not
possible to define the induction principle (and many other functions on
terms) with two mutually recursive definitions, one on terms and
another one on vectors of terms, as follows
(|H1|, |H2|, \ldots are the induction hypotheses):

\begin{Verbatim}[fontsize=\small]
Fixpoint term_rect t : P t :=
  match t as t return P t with
    | Var x => H1 x
    | Fun f v => H2 f (terms_rect (arity f) v)
  end
with terms_rect n (v : terms n) : Q v :=
  match v as v return Q v with
    | Vnil => H3
    | Vcons t' n' v' => H4 (term_rect t') (terms_rect n' v')
  end.
\end{Verbatim}

Instead, we have to use a nested fixpoint:

\begin{Verbatim}[fontsize=\small]
Fixpoint term_rect t : P t :=
  match t as t return P t with
    | Var x => H1 x
    | Fun f v =>
      let fix terms_rect n (v : terms n) : Q v :=
        match v as v return Q v with
          | Vnil => H3
          | Vcons t' n' v' => H4 (term_rect t') (terms_rect n' v')
        end in H2 f (terms_rect (arity f) v)
  end.
\end{Verbatim}

Since inner fixpoints are anonymous, we need to duplicate its
definition to give it a name and prove that both expressions are
indeed equal. Finally, another disadvantage is that definition
unfolding creates big terms.

\subsection{Contexts}

Contexts are defined as terms with a {\em unique} hole in a similar
way (module |AContext|):

\begin{Verbatim}[fontsize=\small]
Notation terms := (vector term).
Inductive context : Type :=
  | Hole : context
  | Cont : forall (f : Sig) (i j : nat), i + S j = arity f ->
    terms i -> context -> terms j -> context.
Fixpoint fill (c : context) (t : term) : term := ...
\end{Verbatim}

Thanks to the use of dependent types, contexts are well-formed by
construction and replacing a hole by some term always leads to a
well-formed term.

As the reader may already have remarked in the previous declarations
which are directly taken from the CoLoR files, Coq offers a mechanism
for automatically inferring, to some extent, the missing arguments and
types. This is a very important feature, especially with polymorphic
and dependent types, that are heavily used in CoLoR. For instance, for
a context, it is sufficient to write |(Cont h ti c tj)| where |h| is
of type |i+Sj=arity f|. Then, Coq can infer that this is in fact
|(Cont Sig f i j h ti c tj)|.

\subsection{Interpretations and substitutions}
\label{sec-int}

We now come to the interpretation of terms into some non-empty domain
$D$ given an interpretation function $I_f:D^n\a D$ for each function
symbol $f$ of arity $n$ (module |AInterpretation|). This gives $D$ a 
structure of a $\S$-algebra where $\S$ is the signature.

\begin{Verbatim}[fontsize=\small]
Definition naryFunction A B n := vector A n -> B.
Definition naryFunction1 A := naryFunction A A.
Record interpretation : Type := mkInterpretation {
  domain :> Type;
  some_elt : domain;
  fint : forall f : Sig, naryFunction1 domain (arity f) }.
\end{Verbatim}
where |domain| is the domain $D$ and |fint| corresponds to the family
of functions $I_f$.
Given a valuation $\r:\cX\a D$ for the variables of $t$ (|valuation|),
the interpretation of a term $t$, written $\I{t}\r$ (|term_int t valuation|),
is the recursive application of the interpretation functions $I_f$:

\begin{Verbatim}[fontsize=\small]
Variable I : interpretation.
Definition valuation := variable -> (domain I).
Variable xint : valuation.
Fixpoint term_int t :=
  match t with
    | Var x => xint x
    | Fun f ts => fint I f (Vmap term_int ts)
  end.
\end{Verbatim}

A substitution is then nothing but an interpretation on the domain of
terms by taking $I_f(t_1,\ldots,t_n)=f(t_1,\ldots,t_n)$.

\begin{Verbatim}[fontsize=\small]
Definition I0 := mkInterpretation (Var 0) (@Fun Sig).
Definition substitution := valuation I0.
Definition sub : substitution -> term -> term := @term_int Sig I0.
\end{Verbatim}

Interpretations play an important role in termination. Indeed, given
an interpretation $I$ and a relation $>$ on the domain $D$ of $I$, the
relation $>_I$ on terms such that $t>_I u$ if $\I{t}\r>\I{u}\r$ for
all valuation $\r:\cX\a D$ for the variables of $t$ and $u$, is
well-founded when $>$ is well-founded \cite{manna70hicss}. The
relation $>_I$ is always stable under substitutions and it is stable under
contexts if the functions $I_f$ are monotone wrt $>$ in every argument
(module |AWFMInterpretation|). We will come back to this point in
Section \ref{sec-manna-ness}.

Before that, we define some basic properties of relations on terms
(module |ARelation|):

\begin{Verbatim}[fontsize=\small]
Definition preserve_vars := forall t u, R t u -> incl (vars u) (vars t).
Definition substitution_closed :=
  forall t1 t2 s, R t1 t2 -> R (sub s t1) (sub s t2).
Definition context_closed :=
  forall t1 t2 c, R t1 t2 -> R (fill c t1) (fill c t2).
Definition rewrite_ordering := substitution_closed /\ context_closed.
\end{Verbatim}

\noindent
where |vars| is the set of variables of a term (module |AVariables|),
modeled using |FSets|: the standard finite sets library of Coq.

\subsection{Rewriting}

Having defined the notions of context and substitutions, we can now
define rewriting (module |ATrs|). Rewrite relations are defined from
sets of rules, a rule simply being a pair of terms:

\begin{Verbatim}[fontsize=\small]
Record rule : Type := mkRule { lhs : term; rhs : term }.
\end{Verbatim}

For a system to terminate, it is necessary that the left-hand side is
not a variable and that all the variables occurring in the right-hand
side also occur in the left hand-side. These conditions are not
required at this stage but will be part of the termination conditions
to check.

The standard rewrite relation generated from a list |R| of rewrite
rules is then defined as follows:

\begin{Verbatim}[fontsize=\small]
Definition red u v := exists l r c s,
  In (mkRule l r) R /\ u = fill c (sub s l) /\ v = fill c (sub s r).
\end{Verbatim}

We have |red : term -> term -> Prop|, so |red| is a relation on
terms. We have a reduction step |red u v|, usually written as $u\a_Rv$,
whenever there exist a term $l$, a term $r$, a context $C$ and a 
substitution $\s$ such that $l\a r\in R$, $u=C[l\s]$ and $v=C[r\s]$.

The transformations done by modern termination techniques in fact lead
to more complex relations. First, rewriting can occur at the top of a
term only or, conversely, never at the top:

\begin{Verbatim}[fontsize=\small]
Definition hd_red u v := exists l r s,
  In (mkRule l r) R /\ u = sub s l /\ v = sub s r.
Definition int_red u v := exists l r c s, c<>Hole /\
  In (mkRule l r) R /\ u = fill c (sub s l) /\ v = fill c (sub s r).
\end{Verbatim}

\noindent
And one may consider (top) rewriting modulo some other relation:

\begin{Verbatim}[fontsize=\small]
Definition red_mod := red E # @ red R.
Definition hd_red_Mod := S @ hd_red R.
Definition hd_red_mod := red E # @ hd_red R.
\end{Verbatim}

\noindent
where |@| denotes the composition of two relations, and |#| the
reflexive and transitive closure of a relation (module |RelUtil|).

\subsection{Termination}\label{subsection:termination}

We now come to the definition of termination itself. In the
introduction, we defined it as the absence of infinite
rewrite sequences (chains), as it is often defined in the literature.

An alternative definition is based on the notion of well-foundedness:
a relation
$R$ is well-founded on a class $X$ if every non-empty subset of $X$
has a minimal element with respect to $R$.

The two definitions are equivalent in classical logic under the Axiom
of Choice.

The interesting aspect of the second definition is that it provides an induction
principle that is a particular case of transfinite induction: given a
relation $S$ that is well-founded on $X$, a property $P$ holds for all
the elements of $X$ if, for all $x,y\in X$, $P(x)$ holds whenever
$P(y)$ holds for all $y$ such that $ySx$.

Such a generic induction principle, with $P$ as a parameter, is defined
in the Coq standard library (module |Wellfounded|). Because its use
requires to write rewrite steps the other way around ($y\,{}_R\!\!\la
x$ instead of $x\a_Ry$), in CoLoR, as it is traditional in rewriting
theory, we prefer to use the following dual definition (module
|SN|):

\begin{Verbatim}[fontsize=\small]
Inductive SN x : Prop := SN_intro : (forall y, R x y -> SN y) -> SN x.
Definition WF := forall x, SN x.
\end{Verbatim}

However, we provide lemmas to go from the CoLoR representation to the
Coq representation:
\begin{Verbatim}[fontsize=\small]
Lemma WF_transp_wf : WF (transp R) -> well_founded R.
Lemma wf_transp_WF : well_founded (transp R) -> WF R.
\end{Verbatim}
where |transp| is a transposition of a relation: |transp R x y = R y x|.

Well-founded relations can also be used to define functions
recursively. In Coq, the well-foundedness proof itself can be used as
an extra-argument to make explicit the decrease of the arguments in
the recursive calls.

Hence, CoLoR can also be seen as a toolbox for defining functions by
well-founded induction and proving the totality of such functions.
Indeed, CoLoR provides various results on the theory of arbitrary
(well-founded) relations (and not only rewrite relations) like, for
instance, the multiset extension of a relation (directory
|Util/Multiset|) and some general results on the union and composition
of well-founded relations (directory |Util/Relation|).

\bigskip

Finally, we formalized the relations between the two definitions of
termination.

When a relation is finitely branching (the set of successors is always
finite, although not necessarily bounded), we proved that the
existence of an infinite chain implies that the relation is not
well-founded (module |IS_NotSN|). Proving this implication for
non-finitely-branching relations would require to develop some general
theory of ordinals. However, we proved that a rewrite relation
generated from a set of rules is finitely branching whenever the set
of rules is finite (module |AReduct|).

For proving the converse, that is, that there is an infinite chain if
the relation is not well-founded (module |NotSN_IS|), as already
mentioned, we need to use the Axiom of Excluded Middle and the Axiom
of (Dependent) Choice:

\begin{Verbatim}[fontsize=\small]
Definition IS R f := forall i, R (f i) (f (S i)).
Definition classic_left_total R := forall x, exists y, R x y.
Axiom dep_choice : forall (B : Type) (b : B) (R : relation B),
  classic_left_total R -> exists f, IS R f.
\end{Verbatim}
where |IS| stands for infinite-sequence and expresses that there
is an infinite reduction in |R|. The |dep_choice| axiom is a
consequence of a more general choice axiom (\emph{cf.}
|DepChoicePrf| in CoLoR and |ClassicalChoice| in Coq
standard library).

\section{Dependency pairs}
\label{sec-dp}

In this section, we explain the formalization and proof of a key
notion of modern termination techniques: the notion of dependency pairs
\cite{
arts00tcs,hirokawa05ic,giesl06jar}.

We call a function symbol $f$ \emph{defined} if there is a rule which
left-hand side is headed by $f$; otherwise it is a \emph{constructor}.
Further denote by $\atop~$ (resp. $\aint~$) rewriting at the top
(resp. below the top); defined in Coq as |hd_red| (resp. |int_red|).

Let $R$ be a set of rewrite rules, and assume that the rewrite
relation generated by $R$ does not terminate. Let $\cT_\inf$ be the
(non-empty) set of non-terminating terms whose subterms are all
terminating. Then, for all $t\in\cT_\inf$, there is a rule $l\a r\in
R$ and a subterm $s$ of $r$ headed by a defined symbol
that is not a strict subterm of $l$ \cite{dershowitz04iclp}, and such
that $t\aint{}_R^*~ l\s\atop_R r\s\tge s\s\in\cT_\inf$.

The pairs $l\a s$ such that $l\a r\in R$ and $s$ is a subterm of $r$
headed by a defined symbol that is not a strict subterm of $l$ are
called the {\em dependency pairs} of $R$, and the relation
$\aint{}_R^*\atop_R$ (relation composition is written by
juxtaposition) is called a {\em chain} (module |ADP|):

\begin{Verbatim}[fontsize=\small]
Fixpoint mkdp (S : rules) : rules :=
  match S with
    | nil => nil
    | a :: S' => let (l,r) := a in
      map (mkRule l) (filter (negb_subterm l) (calls R r)) ++ mkdp S'
  end.
Definition dp := mkdp R.
Definition chain := int_red R # @ hd_red dp.
\end{Verbatim}

\noindent
where |++| (concatenation), |map| and |filter| are usual functions of
the Coq standard library on lists, |negb_subterm l| says if a term is
not a subterm of |l| (|negb| standing for negation on the |bool| type,
as opposed to |not| on |Prop|), and |calls R r| gives the list of subterms of
|r| which are headed by a symbol defined by a rule of |R|.

In the literature, dependency pairs are generally expressed in the
extended signature $\S=\S\cup\{f^\sharp\mid f\in\S\}$ by replacing in
both sides of a dependency pair the top symbol by their $\sharp$
version. As a consequence, $\aint{}_R$ can be replaced by $\a_R$ since
no rule of $R$ can be applied at the top of a term of the form
$f^\sharp(t_1, \ldots, t_n)$. This transformation is also formalized in CoLoR in the
module |ADuplicateSymb|.

Then, the main theorem of dependency pairs says (module |ADP|):

\begin{Verbatim}[fontsize=\small]
Variable hyp1 : forallb (@is_notvar_lhs Sig) R = true.
Variable hyp2 : rules_preserve_vars R.
Lemma WF_chain : WF chain -> WF (red R).
\end{Verbatim}

As just explained, the classical proof consists of assuming an
infinite sequence of |R| steps and showing that one can build an infinite
sequence of |chain| steps. But a direct constructive proof can be
given by well-founded induction on |chain|. As shown in
\cite{blanqui06wst-hodp}, the proof technique is the same as the one
based on Tait and Girard computability/reducibility predicates
\cite{girard88book} for proving the termination of $\b$-reduction, the
correctness of the notion of computability closure
\cite{
blanqui07jpbirthday}, or the
termination of the higher-order recursive path ordering
\cite{jouannaud99lics}.

First, we proceed by well-founded induction on |chain|. Second, we
prove that every term $t$ terminates by induction on $t$ and
well-founded induction on the terminating subterms of $t$. If $t$ is a
variable, then it follows from the assumption |hyp1| that no rule
left-hand side is a variable (module |ASN|). Otherwise, by definition
of termination, it suffices to prove that every reduct of $t$
terminates. If the reduction does not occur at the top, then the
conclusion follows from the induction hypotheses. Otherwise, there is
a rule $l\a r$ and a substitution $\s$ such that $t=l\s$ and the
reduct is $r\s$. Then, we can conclude by proving the following two
lemmas:

\begin{enumerate}
\item
Every subterm of $r\s$ headed by a defined symbol terminates, either
because it is a strict subterm of $l\s$, or because it is a |chain|
reduct. In both cases, we can conclude by induction hypothesis.
\item
We have $r\s=s\t$ where $s$ is the constructor term obtained from
$r\s$ by replacing subterms headed by a defined symbol by distinct
fresh variables, and $\t$ is the substitution mapping each one of
these fresh variables to the corresponding subterms. Then, one can
prove that, for all constructor terms $s$ and terminating
substitutions $\t$ (\ie $x\t$ terminates for all $x$), $s\t$
terminates (module |ASN|).
\end{enumerate}

The term $s$ is generally called the (constructor) {\em cap} of $r\s$,
and the substitution $\t$ the corresponding {\em alien} substitution
of $t$. The cap and the alien substitution are unique up to the
renaming of fresh variables used in the cap. The definitions of cap
and aliens provide a nice example of higher-order, dependent and
polymorphic function (module |ACap|). It uses an auxiliary function
|capa| which, for every term $t$, computes an element of type:

\begin{Verbatim}[fontsize=\small]
Definition Cap := { k : nat & (terms k -> term) * terms k }.
\end{Verbatim}

\noindent
which is a dependent triple $(k,f,v)$ where:

\begin{itemize}
\item
$k$ is the number of aliens of $t$ (function |nb_aliens|),
\item
$f$ is a function which, given a vector $x$ of $k$ terms, returns the
  cap with the $i$-th alien replaced by the $i$-th term of $x$
  (function |fcap|),
\item
$v$ is the vector of the $k$ aliens of $t$ (function |aliens|).
\end{itemize}

The function |capa| is then defined as follows:

\begin{Verbatim}[fontsize=\small]
Fixpoint capa (t : term) : Cap :=
  match t with
    | Var x => mkCap (fun _ => t, Vnil)
    | Fun f ts =>
      if defined f R then
	mkCap (fun v => Vnth v (lt_O_Sn 0), Vcons t Vnil)
      else
	let cs := Vmap capa ts in
	mkCap (fun v => Fun f (Vmap_sum cs v), conc cs)
  end.
\end{Verbatim}

\noindent
where |mkCap| is a function to construct term of type |Cap|,
|lt_0_Sn| is a proof that |0 < S n| for arbitrary |n|
and |conc| concatenates all the aliens of a vector of caps:

\begin{Verbatim}[fontsize=\small]
Fixpoint conc n (cs : Caps n) : terms (sum cs) :=
  match cs as cs return terms (sum cs) with
    | Vnil => Vnil
    | Vcons c _ cs' => Vapp (aliens c) (conc cs')
  end.
\end{Verbatim}

\noindent
and, given a vector |cs| of caps and a vector |v| of |(sum cs)| terms,
|Vmap_sum| breaks |v| in vectors which sizes are the numbers of aliens
of every cap of |cs|, applies every |fcap| to the corresponding
vector, and concatenates all the results:

\begin{Verbatim}[fontsize=\small]
Fixpoint Vmap_sum n (cs : Caps n) : terms (sum cs) -> terms n :=
  match cs as cs in vector _ n return terms (sum cs) -> terms n with
    | Vnil => fun _ => Vnil
    | Vcons c _ cs' => fun ts =>
      let (hd,tl) := Vbreak ts in Vcons (fcap c hd) (Vmap_sum cs' tl)
  end.
\end{Verbatim}

Finally, the cap and aliens are obtained as follows:

\begin{Verbatim}[fontsize=\small]
Definition cap t := match capa t with existS n (f,_) => f (fresh_for t n) end.
Definition alien_sub t := fsub (maxvar t) (aliens (capa t)).
\end{Verbatim}
where |existS x P| is the single constructor of the subset type behind
the notation \verb#{x | P}#, |fresh_for t n| is a vector of |n|
variables fresh for |t| and |fsub x0 n v| is the substitution
$\{x_0+1\mapsto v_1, \ldots, x_0+n\mapsto v_n\}$.

\section{Dependency graph decomposition}
\label{sec-graph}

The next major result of the dependency pair framework is based on the
analysis of the possible sequences of function calls. This is accomplished
by means of an analysis of the so-called \emph{dependency graph}, $\cG(R)$.
The dependency graph is the graph where nodes are the
dependency pairs of $R$ and edges are given by the relation
$\aint{}_R^*$: there is an edge from $l_1\a s_1$ to $l_2\a s_2$ if
$s_1\s_1{\aint{}_R^*}~l_2\s_2$ for some substitutions $\s_1$ and
$\s_2$. Then, the relation $\aint{}_R^*\atop_R$ terminates if, for
every strongly connected component $C$ of $\cG(R)$,
$\aint{}_C^*\atop_C$ terminates \cite{giesl02jsc}. This important
result allows one to split a termination problem into simpler
ones that can be dealt with independently (and in
parallel) using different techniques.

In CoLoR, $\cG(R)$ is formalized as an instance, taking |(int_red R#)|
for |S|, and |(dp R)| for |D|, of the following relation
|hd_rules_graph| on rules, where |shift| |p| is the substitution
mapping a variable |x| to the variable |x+p| (using a shift and a
substitution is equivalent and no more difficult than using two
substitutions):

\begin{Verbatim}[fontsize=\small]
Variable S : relation term.
Variable D : rules.
Definition hd_rules_graph a1 a2 := In a1 D /\ In a2 D
  /\ exists p, exists s, S (sub s (rhs a1)) (sub s (shift p (lhs a2))).
\end{Verbatim}

However, the graph $\cG(R)$ is generally not decidable. To address
this problem, various decidable over-approximations have been
introduced. The most common one is based on unification (module
|AUnif|): there is an edge from $l_1\a s_1$ to $l_2\a s_2$ if
$\rencap(s_1)$ and $l_2$ are unifiable, where $\rencap(s_1)$ (module
|ARenCap|) is like the cap of $s_1$ defined in the previous section
but with variables considered as aliens too and alien subterms
replaced by fresh variables not occurring in $l_2$
\cite{arts00tcs}. Note that, by definition of the cap, two occurrences
of the same alien subterm are replaced by distinct fresh
variables. Hence, $\rencap(s_1)$ is linear and its variables are
distinct from those of $l_2$. The Coq formalization (module |ADPUnif|)
is as follows (|<<| is the notation for relation inclusion):

\begin{Verbatim}[fontsize=\small]
Variables R D : rules.
Definition connectable u v := unifiable (ren_cap R (S (maxvar v)) u) v.
Definition dpg_unif (r1 r2 : rule) :=
  In r1 D /\ In r2 D /\ connectable (rhs r1) (lhs r2).
Lemma dpg_unif_correct : hd_rules_graph (red R #) D << dpg_unif.
\end{Verbatim}

Now, to avoid the computation inside Coq of the strongly connected
components of the chosen over-approximation of $\cG(R)$ (represented
hereafter by a boolean function |approx|) and of their topological
ordering, and also to allow the user to deal with various components
at the same time, we introduce a notion of valid decomposition of the
set of dependency pairs:

\begin{Verbatim}[fontsize=\small]
Variable approx : rule -> rule -> bool.
Notation decomp := (list rules).
Fixpoint valid_decomp (cs : decomp) : bool :=
  match cs with
    | nil => true
    | ci :: cs' => valid_decomp cs' &&
      forallb (fun b =>
        forallb (fun cj =>
          forallb (fun c => negb (approx b c)) cj)
        cs')
      ci
  end.
\end{Verbatim}

A decomposition $(c_1,\ldots,c_n)$ (the order is important) is valid
if for all $i$, for every rule $b$ in $c_i$, for all $j>i$, and for every
rule $c$ in $c_j$, there is no edge from $b$ to $c$. Hence, we can
proceed by induction on the size $n$ of the decomposition to prove the
decomposition theorem under the following assumptions (module
|ADecomp|):

\begin{Verbatim}[fontsize=\small]
Definition Graph x y := approx x y = true.
Variable approx_correct : hd_rules_graph S D << Graph.
Lemma WF_decomp :
  forall (hypD : rules_preserve_vars D) (cs : decomp) (hyp1 : incl D (flat cs))
    (hyp2 : incl (flat cs) D) (hyp3 : valid_decomp cs = true)
    (hyp4 : lforall (fun ci => WF (hd_red_Mod S ci)) cs),
    WF (hd_red_Mod S D).
\end{Verbatim}

Indeed, if $R_1$ and $R_2$ are two terminating relations, then
$R_1\cup R_2$ terminates whenever $R_1R_2\sle R_2R_1$ (module
|Union|), which is trivially the case here since $R_1R_2$ is empty
(there is no edge from $c_i$ to $c_j$ if $j>i$).

\section{Reduction pairs}
\label{sec-manna-ness}

A general termination technique consists of checking that every rule
is included in some {\em reduction ordering}, \ie a well-founded
rewrite relation, like, for instance, the recursive path ordering
\cite{dershowitz82tcs} (module |AMannaNess|):

\begin{Verbatim}[fontsize=\small]
Variables (R : rules) (succ : relation term).
Definition reduction_ordering := WF R /\ rewrite_ordering R.
Definition compat := forall l r : term, In (mkRule l r) R -> succ l r.
Lemma manna_ness : reduction_ordering succ -> compat -> WF (red R).
\end{Verbatim}

This basic result can be extended to (top) rewriting modulo by using
(weak) {\em reduction pairs} \cite{kusakari99ppdp}, that is, pairs of
relations $(\succ,\succeq)$ closed by substitution such that $\succ$
is well-founded, ${\succeq \cdot \succ}\sle{\succ}$ and $\succeq$ (and
$\succ$) are closed by context (module |ARelation|):

\begin{Verbatim}[fontsize=\small]
Definition absorb A (R S : relation A) := S @ R << R.
Record Weak_reduction_pair : Type := mkWeak_reduction_pair {
  wp_succ : relation term;
  wp_succ_eq : relation term;
  wp_subs : substitution_closed wp_succ;
  wp_subs_eq : substitution_closed wp_succ_eq;
  wp_cont_eq : context_closed wp_succ_eq;
  wp_absorb : absorb wp_succ wp_succ_eq;
  wp_succ_wf : WF wp_succ }.
Lemma manna_ness_hd_mod : forall wp : Weak_reduction_pair Sig,
  compat (wp_succ_eq wp) E -> compat (wp_succ wp) R -> WF (hd_red_mod E R).
\end{Verbatim}

More generally, if all rules are included in $\succeq$, then all rules
included in $\succ$ can be removed \cite{endrullis08jar}. Reduction
pairs can therefore be used to simplify termination problems. This is
formalized in CoLoR by a functor
\cite{chrzaszcz03tphol} 
taking as
argument boolean functions representing some decidable
under-approximations of the relations $\succ$ and $\succeq$ (module
|ARedPair|):

\begin{Verbatim}[fontsize=\small]
Module Type WeakRedPair.
  Parameter Sig : Signature. Notation term := (@term Sig).
  Parameter succ : relation term.
  Parameter wf_succ : WF succ.
  Parameter sc_succ : substitution_closed succ.
  Parameter bsucc : term -> term -> bool.
  Parameter bsucc_sub : rel bsucc << succ.
  ...
End WeakRedPair.
\end{Verbatim}

\begin{Verbatim}[fontsize=\small]
Module WeakRedPairProps (Import WP : WeakRedPair).
  Notation rule := (rule Sig). Notation rules := (rules Sig).
  Definition wp := mkWeak_reduction_pair
    sc_succ sc_succeq cc_succeq succ_succeq_compat wf_succ.
  Lemma WF_wp_hd_red_mod : forall E R,
    forallb (brule bsucceq) E = true ->
    forallb (brule bsucceq) R = true ->
    WF (hd_red_mod E (filter (brule (neg bsucc)) R)) ->
    WF (hd_red_mod E R).
End WeakRedPairProps.
\end{Verbatim}

This functor also provides high-level tactics \cite{delahaye00lpar}
for applying its lemmas and automatically checking their conditions:

\begin{Verbatim}[fontsize=\small]
  Ltac check_eq := vm_compute; refl.
  Ltac do_prove_termination prove_cc_succ lemma := apply lemma;
    match goal with
    | |- context_closed _ => prove_cc_succ
    | |- WF _ => idtac
    | |- _ = _ => check_eq || fail "some rule is not in the ordering"
    end.
  Ltac prove_termination prove_cc_succ :=
    let prove := do_prove_termination prove_cc_succ in
    match goal with
    | |- WF (red _) => prove WF_rp_red
    | |- WF (red_mod _ _) => prove WF_rp_red_mod
    | |- WF (hd_red_mod _ _) => prove WF_wp_hd_red_mod ...
    end.
\end{Verbatim}

The first tactic tries to prove a goal of the form |_ = _| by first
reducing each side of the equality to their normal form using Coq
efficient normalization procedure |vm_compute|
\cite{gregoire02icfp}, 
and then by checking the
syntactic equality of the resulting terms.

The second tactic takes as argument another tactic |prove_cc_succ| for
checking that |succ| is closed by context, and the termination |lemma|
to use. The generated subgoals of the form |WF _| are left unchanged
(|idtac|), and the generated subgoals of the form |_ = _| are proved
by using the first tactic.

Finally, the third tactic takes |prove_cc_succ| as argument and,
depending on the form of the goal, calls the second tactic with the
appropriate termination lemma (\eg |WF_wp_hd_red_mod| for a relative
top termination problem).

\bigskip

There are various ways to build reduction orderings/pairs. As already
mentioned in Section \ref{sec-int}, interpretation in some
well-founded domain is a popular one \cite{manna70hicss}.

Indeed, an interpretation |I| and a relation |R| on the domain of |I|
provide a relation on terms by universally quantifying on all possible
interpretations of variables (module |AWFMInterpretation|):

\begin{Verbatim}[fontsize=\small]
Definition IR : relation term :=
  fun t u => forall xint, R (term_int xint t) (term_int xint u).
\end{Verbatim}

Many properties satisfied by |R| are also satisfied by |IR|, in
particular well-foundedness. Moreover, |IR| is closed by
substitution. However, for |IR| to be closed by context, the
interpretation of each symbol needs to be monotone wrt |R|:

\begin{Verbatim}[fontsize=\small]
Definition monotone := forall f, Vmonotone1 (fint I f) R.
Lemma IR_reduction_ordering : monotone -> WF R -> reduction_ordering IR.
\end{Verbatim}

Based on the notions of a reduction pair and of interpretation
into some well-founded domain, a generic module (|AMonAlg|) of 
(extended) monotone algebras is defined, following the presentation 
of \cite{endrullis08jar}, with support for total, relative and 
relative-top termination.

\section{Polynomial interpretations}
\label{sec-poly}

In this section, we present a formalization of a widely used class of
interpretations on the well-founded domain of natural numbers:
polynomial interpretations \cite{lankford79tr,contejean05jar}.

Our current formalization of integer polynomials (module |Polynom|) is
simple, but sufficient for our purpose since polynomials
used in termination proofs are often small (degree and coefficients
are often bounded by small constants like 1 or 2).

The type of polynomials depends on the maximum number $n$
of variables. A polynomial is represented by a list of pairs made of
an integer and a monomial, a monomial being a vector of size $n$ made
of the powers of each variable:

\begin{Verbatim}[fontsize=\small]
Notation monom := (vector nat).
Definition poly n := list (Z * monom n).
\end{Verbatim}

For instance, if $n=2$ and the variables are denoted by
$x_1,\ldots,x_n$, then the monomial $x_1^3x_2$ is represented by the
vector $(3,1)$.

This representation is not canonical: a polynomial can be represented
in various ways. It is however easy to compute the coefficient of a
monomial:

\begin{Verbatim}[fontsize=\small]
Fixpoint coef n (m : monom n) (p : poly n) : Z :=
  match p with
    | nil => 0
    | cons (c,m') p' =>
      match monom_eq_dec m m' with
	| left _ => c + coef m p'
	| right _ => coef m p'
      end
  end.
\end{Verbatim}

The module |Polynom| then provides basic operations on polynomials:
addition, subtraction, multiplication, power, composition and
evaluation to an integer (|Z|) given values for variables.

Now, a polynomial interpretation consists of associating to every
function symbol of arity $n$, an integer polynomial with $n$ variables
(module |APolyInt|).

\begin{Verbatim}[fontsize=\small]
Definition PolyInterpretation := forall f : Sig, poly (arity f).
\end{Verbatim}

Then, every term with $n$ variables can be interpreted by an integer
polynomial with $n$ variables, by recursively composing the
polynomials interpreting the function symbols occurring in the
term. However, to define this polynomial, we need to know the number
of variables in advance. To this end, we use an intermediate
representation for terms where variables (which are represented by
natural numbers) are bounded (module |ABterm|):

\begin{Verbatim}[fontsize=\small]
Variable k : nat.
Inductive bterm : Type :=
  | BVar : forall x : nat, x<=k -> bterm
  | BFun : forall f : Sig, vector bterm (arity f) -> bterm.

Fixpoint inject_term (t : term) : maxvar_le k t -> bterm := ...

Variable PI : PolyInterpretation.
Fixpoint termpoly k (t : bterm k) : poly (S k) :=
  match t with
    | BVar x H => ((1)%Z, mxi (gt_le_S (le_lt_n_Sm H))) :: nil
    | BFun f v => pcomp (PI f) (Vmap (@termpoly k) v)
  end.
\end{Verbatim}
where |mxi H| is the monomial $x_i$ if $H$ is a proof of $i < n$.

Now, for proving termination of some rewrite system, the
polynomial interpretation must satisfy two conditions:
\begin{itemize}
  \item polynomials must be monotone in each variable;
  \item for every rule $l\a r$, the interpretation of $l$ 
    (polynomial $P_l$), must be strictly bigger than the 
    interpretation of $r$ (polynomial $P_r$), for all 
    evaluations of variables in $\bN$.
\end{itemize}

The latter problem corresponds directly to the positiveness
test for the polynomial $P_l-P_r-1 \ge 0$, which is undecidable in 
general. We follow the usual approach taken in most termination 
provers and use absolute positiveness check for that purpose
\ie, a polynomial is absolutely positive if all its coefficients 
are non-negative \cite{contejean05jar}.

We also use a simple test for (strict) monotonicity, by testing 
that each monomial $x_i$ is (strictly) positive.
However, we plan to implement more general
conditions, especially for polynomials of degree two.

\begin{Verbatim}[fontsize=\small]
Program Definition rulePoly_ge rule := 
  let l := lhs rule in 
  let r := rhs rule in
  let m := max (maxvar l) (maxvar r) in
  termpoly (@inject_term _ m l _) - termpoly (@inject_term _ m r _).

Definition rulePoly_gt rule := rulePoly_ge rule - 1.
\end{Verbatim}

\noindent The minus ($-$) operator and the constant $1$ above are overwritten 
notations for polynomial operations.

Then, given a polynomial interpretation, one can define an instance
of a monotone algebra (|AMonAlg|), as described in the preceding
section, and use the associated tactics to simplify termination 
problems, by removing strictly decreasing rules.

\section{Example of automatically generated termination proof}
\label{sec-example}

In this section, we present {\sl in extenso} an example of Coq script
automatically generated by Rainbow from some simple termination
certificate.

The rewrite system that we consider is:

\begin{rewc}
minus(x,zero) & x\\
minus(succ(x),succ(y)) & minus(x,y)\\
quot(zero,succ(y)) & zero\\
quot(succ(x),succ(y)) & succ(quot(minus(x,y),succ(y)))\\
\end{rewc}

This system computes the quotient of two natural numbers encoded in
unary notation. This system is not simply terminating (by
taking $y=succ(x)$, the right-hand side of the fourth rule can be
embedded in the corresponding left-hand side). It can however be dealt
with by a simplification ordering \cite{dershowitz90book} after
applying some argument filtering \cite{arts00tcs} (by erasing the
first argument of $minus$), another simple but very useful technique
that is formalized in the module |AFilter| but that we will not
explain in this paper. Hence, we do the argument filtering by hand and
consider the resulting system instead:

\begin{rewc}
minus(zero) & zero\\
minus(succ(x)) & succ(minus(x))\\
quot(zero,succ(y)) & zero\\
quot(succ(x),succ(y)) & succ(quot(minus(x),succ(y)))\\
\end{rewc}

\noindent
and the following termination certificate (in the Rainbow format 
with some XML tags removed for the sake of simplicity; note that
Rainbow also supports the CPF format used by many termination 
provers) that could be automatically generated by some termination 
prover:

\begin{Verbatim}[fontsize=\small]
<dp>
 <decomp><graph><hde/></graph>
  <component>
   <rules><!-- minus(succ(x)) -> minus(x) -->...</rules>
   <manna_ness>
    <poly_int>
     <fun>zero</fun>
     <polynomial><!-- 0 -->...</polynomial>
     <fun>succ</fun>
     <polynomial><!-- 1.x_1 + 2 -->...</polynomial>
     <fun>minus</fun>
     <polynomial><!-- 1.x_1 + 1 -->...</polynomial>
     <fun>quot</fun>
     <polynomial><!-- 1.x_1.x_2 + 1.x_1 + 1.x_2 -->...</polynomial>
    </poly_int>
    <trivial/>
   </manna_ness>
  </component>
  <component>
   <rules><!-- quot(succ(x),succ(y)) -> minus(x) -->...</rules>
  </component>
  <component>
   <rules><!--quot(succ(x),succ(y)) -> quot(minus(x),succ(y))-->...</rules>
   <manna_ness>
    <poly_int>
     <fun>zero</fun>
     <polynomial><!-- 0 -->...</polynomial>
     <fun>succ</fun>
     <polynomial><!-- 1.x_1 + 2 -->...</polynomial>
     <fun>minus</fun>
     <polynomial><!-- 1.x_1 + 1 -->...</polynomial>
     <fun>quot</fun>
     <polynomial><!-- 1.x_1.x_2 + 1.x_1 + 1.x_2 -->...</polynomial>
    </poly_int>
    <trivial/>
   </manna_ness>
  </component>
 </decomp>
</dp>
\end{Verbatim}

This certificate proposes to prove the termination of the system by
the following procedure:

\begin{enumerate}
\item
apply the dependency pair transformation,
\item
decompose the resulting problem into 3 components:

\begin{enumerate}[(w)]
\item
eliminate all the rules of the first component that strictly decrease
in the given polynomial interpretation, resulting in an empty set of
rules (XML tag {\small|<trivial/>|}),

\item
do nothing with the second component since it contains no loop in the
dependency graph,

\item
eliminate all the rules of the third component that strictly decrease
in the given polynomial interpretation, resulting in an empty set of
rules (XML tag {\small|<trivial/>|}).
\end{enumerate}
\end{enumerate}

The Coq script generated by Rainbow is the following. It first
defines the signature and the set of rules:

\begin{Verbatim}[fontsize=\small]
Require Import (* necessary CoLoR modules *) ...
Open Scope nat_scope.

(* set of function symbols *)
Module M. Inductive symb : Type := minus : symb | ... End M.
Definition beq_symb (f g : M.symb) : bool :=
  match f, g with M.minus, M.minus => true | ... | _, _ => false end.
Lemma beq_symb_ok : forall f g : M.symb, beq_symb f g = true <-> f = g.
  Proof. beq_symb_ok. Qed.
Definition ar (s : M.symb) : nat := match s with M.minus => 1 | ... end.

(* signature 1 *)
Module S1.
  Definition Sig := ASignature.mkSignature ar beq_symb_ok.
  Definition Fs : list Sig := M.zero::M.succ::M.quot::M.minus::nil.
  Lemma Fs_ok : forall f : Sig, In f Fs. Proof. list_ok. Qed.
  Definition some_symbol : Sig := M.minus.
  Lemma arity_some_symbol : arity some_symbol > 0. Proof. check_gt. Qed.
End S1.

(* rewrite rules *)
Definition E : ATrs.rules S1.Sig := nil.
Definition R : ATrs.rules S1.Sig := @ATrs.mkRule S1.Sig
  (@Fun S1.Sig M.minus (Vcons (@Fun S1.Sig M.zero Vnil) Vnil))
  (@Fun S1.Sig M.zero Vnil) :: ... :: nil.
Definition rel := ATrs.red_mod E R.
\end{Verbatim}

Then, it defines each parameter required in the termination proof:

\begin{Verbatim}[fontsize=\small]
(* graph decomposition 1 *)
Definition cs1 : list (list (@ATrs.rule S1.Sig)) :=
   (@ATrs.mkRule S1.Sig (@Fun S1.Sig M.minus
       (Vcons (@Fun S1.Sig M.succ (Vcons (@Var S1.Sig 0) Vnil)) Vnil))
     (@Fun S1.Sig M.minus (Vcons (@Var S1.Sig 0) Vnil))
   :: nil) :: ... :: nil.

(* polynomial interpretation 1 *)
Module PIS1.
  Definition sig := S1.Sig.
  Definition trsInt (f : S1.Sig) :=
    match f as f return poly (@ASignature.arity sig f) with
      | M.zero => (0%Z, Vnil) :: nil | ... end.
  Lemma trsInt_wm : PolyWeakMonotone trsInt.
    Proof. PolyWeakMonotone S1.Fs S1.Fs_ok. Qed.
End PIS1.
Module PI1 := PolyInt PIS1.

(* reduction ordering 1 *)
Module WP1 := WP_MonAlg PI1.MonotoneAlgebra.
Module WPR1.
  Include (WeakRedPairProps WP1).
  Ltac prove_cc := PI1.prove_cc_succ_by_refl S1.Fs S1.Fs_ok.
  Ltac prove_termin := prove_termination prove_cc.
End WPR1.

(* polynomial interpretation 2 *) ...
(* reduction ordering 2 *) ...
\end{Verbatim}

Finally, it generates a short proof script for the termination
theorem, each termination technique used in the certificate giving
rise to one tactic call:

\begin{Verbatim}[fontsize=\small]
(* termination proof *)
Lemma termination : WF rel.
Proof.
unfold rel.
dp_trans.
let D := fresh "D" in set_rules_to D;
graph_decomp S1.Sig (hde_bool D) cs1; subst D. 
hde_bool_correct.
right. WPR1.prove_termin.
termination_trivial.
left. co_scc.
right. WPR2.prove_termin.
termination_trivial.
Qed.
\end{Verbatim}

\noindent
|hde_bool| is the over-approximation based on the equality of top
symbols, and |co_scc| is the tactic taking care of acyclic components.

\section{Conclusion}
\label{sec-conclu}

In this paper we presented the outline of our formalization of the
theory of well-founded rewrite relations~\cite{dershowitz90book,terese03book}
in the proof assistant Coq~\cite{coq}. This includes some key notions
(dependency pairs, dependency graph decomposition and reduction pairs)
that are at the heart of modern state-of-the-art automated termination
provers \cite{giesl06jar,hirokawa07ic}.

We also showed how this formalization is successfully used in the
termination competition \cite{tc} for automatically verifying
correctness of termination certificates generated by those automated
termination provers \cite{cpf}.

We think that this work and the related approaches described in Section
\ref{sec-related} should allow the safe use of external termination
provers in proof assistants, and the development of safe proof
assistants where functions and predicates can be defined by rewrite
rules
\cite{dowek03jar,
blanqui05mscs,
walukiewicz08lmcs,dedukti},
which is one of the solutions proposed so far to increase usability of
dependent types in proof assistants. The other, complementary, solution
is the integration of certified decision procedures
\cite{blanqui07csl,blanqui08tcs,strub10csl,coqmt}.

Other termination techniques have already been formalized in CoLoR
that are not described here: argument filtering (directory |Filter|)
\cite{arts00tcs}, multiset ordering (directory |Util/Multiset|),
higher-order recursive path ordering (directory |HORPO|)
\cite{jouannaud99lics,koprowski09aaecc,koprowski08phd}, matrix and
arctic interpretations (directory |MatrixInt|)
\cite{koprowski08rta,koprowski08sofsem}, semantic and root labelling
\cite{zantema95fi,sternagel08rta} (directory |SemLab|), loops
(directory |NonTermin|), subterm criterion (directory |SubtermCrit|)
\cite{hirokawa07ic}, and usable rules (directory |DP|)
\cite{arts00tcs,giesl03lpar,hirokawa07ic}.

The formalizations of the subterm criterion and usable rules are interesting
since their classical proofs do not seem convertible into direct
constructive proofs. Their formalization in Coq therefore requires the
use of classical logic and the Axiom of Choice. It will then be
interesting to compare it with the same development in Isabelle/HOL
\cite{sternagel10rta}.

We also started to formalize Rainbow itself in Coq in order to get an
efficient standalone certificate checker by using Coq extraction
mechanism \cite{letouzey02types}.

\bigskip

{\bf Acknowledgments.} The authors would like to thank very much all
the contributors to the CoLoR library: Julien Bureaux, Solange
Coupet-Grimal, William Delobel, L\'eo Ducas, S\'ebastien Hinderer,
St\'ephane Le Roux, Sidi Ould Biha, Pierre-Yves Strub, Johannes
Waldmann, Qian Wang, Hans Zantema and Lianyi Zhang, and those who
helped us adapt CoLoR to new versions of Coq: Jean-Marc Notin and Eli
Soubiran.

\small
\bibliographystyle{authordate1}

\end{document}